\documentstyle[preprint,prl,oldlfont,aps,epsf]{revtex}
\begin{document}
\title{Folding and Misfolding of Designed Heteropolymer Chains 
       with Mutations.}

\author{G. Tiana$^{1,2}$, R.A. Broglia$^{1,2,3}$,
H.E. Roman$^{1,4}$, E. Vigezzi$^{1,2}$ and E. Shakhnovich$^5$}

\address{$^1$Dipartimento di Fisica, Universit\`a di Milano,
Via Celoria 16, I-20133 Milano, Italy.}

\address{$^2$INFN, Sezione di Milano, Via Celoria 16, I-20133 Milano, Italy.}

\address{$^3$The Niels Bohr Institute, University of Copenhagen,
2100 Copenhagen, Denmark.}

\address{$^4$Institut f\"ur Theoretische Physik III, Universit\"at Giessen,
Heinrich-Buff-Ring 16, 35392 Giessen, Germany.}

\address{$^5$Department of Chemistry, Harvard University, 12 Oxford Street,
Cambridge, MA 02138}

\date{\today}

\maketitle

\smallskip

\leftline{Pacs numbers: 87.15.Da, 61.43.-j, 64.60.Cn, 64.60.Kw}

\begin{abstract}
We study the impact of mutations (changes in amino acid sequence)
on the thermodynamics of simple protein-like heteropolymers consisting
of $N$ monomers, representing the amino acid sequence.
The sequence is designed to fold into its native conformation
on a cubic lattice.
It is found that quite a large fraction, between one half and one third
of the substitutions, which we  call 'cold errors', make important
contributions to the dynamics of the folding process, increasing folding times
typically by a factor of two, the altered chain still folding into the
native structure. Few mutations ('hot errors'), have quite dramatic
effects, leading to protein misfolding. Our analysis reveals that
mutations affect primarily the energetics of the native conformation
and to a much lesser extent the ensemble of unfolded conformations,
corroborating the utility of the ``energy gap'' concept
for the analysis of folding properties of protein-like heteropolymers.
\end{abstract}

\newpage
\narrowtext
The use of genetically altered proteins has become an important experimental 
tool for studying the physical principles that determine the thermodynamical
stability and folding mechanisms of proteins \cite{Fersht}. 
Many experiments suggest that the effect of  mutations, in most cases
is specific to the location and identity of the mutated amino acid,
consistent with the view that mutations most strongly affect
the properties of the native structure. This conclusion was made
also in the theoretical analysis of the statistical mechanics
of heteropolymers representing simple physical models of protein.
It was argued in \cite{NATUR,SG93} that the basic properties of
the denatured state ensemble are self-averaging, i.e. they depend
on amino acid composition rather than on specific sequences. 
In contrast,
the energetics of the native state are very sensitive to specific details of
a sequence.
An alternative view was suggested that mutations may affect
rather the denatured state, if the latter has elements of structure
\cite{Shortle}.
Of central importance in
this quest is the identification of the role individual amino acid residues
play in the cooperative phenomenon of folding (cf. e.g. Ref. \cite{Creighton} and 
refs. therein). 

A useful theoretical approach to study protein folding
is a simplified lattice model. Such models have proved helpful, not
only for elucidating generic features of 
folding proteins \cite{Lat_Mod}, but also    
for the evaluation of specific nucleation sites
for folding kinetics \cite{NUCLEUS,CONSERV}.
The goal of this Letter is to elucidate, for the same model, the 
impact of mutations on protein {\em thermodynamic stability}.

To this end, the thermodynamics  of protein-like heteropolymer chains consisting
of $N$ monomers, representing the amino acid sequence, is studied on a
cubic lattice for modeling folding 
of proteins to their native structure.
The sequence is designed to fold into the 
native conformation, shown in Fig. 1;
this conformation and the sequence are identical
to that previously studied in \cite{NUCLEUS,TRAPS}.
Mutations in the designed amino acid sequence are introduced by 
replacing a single monomer with a different one.

The configurational energy of the chain is given by
\begin{equation}
E = {1\over2}\sum_{i,j}^N U_{m(i),m(j)}~\Delta(\vert {\bf r}_i-{\bf r}_j\vert),
\label{hami}
\end{equation}
where $U_{m(i),m(j)}$ is the 
effective interaction 
potential between monomers 
$m(i)$ and $m(j)$,
${\bf r}_i$ and ${\bf r}_j$ denote their 
lattice positions and 
$\Delta(x)$ is the contact 
function. 
In Eq. (\ref{hami}), the pairwise interaction 
is different from zero when $i$ and $j$ 
occupy nearest-neighbor sites, i.e. $\Delta(1)=1$ 
and $\Delta(a)=0$ for $a>1$.
In addition to these interactions, it is assumed 
that on-site repulsive forces 
prevent two amino acids to occupy the same 
site simultaneously, so that 
$\Delta(0)=\infty$.

In what follows we shall consider 
a twenty-letter representation 
of protein sequences
where $U$ is a 20 $\times$ 20 matrix. This matrix is
taken from the work \cite{MJ} (Table VI) where it was derived
from frequencies of contacts between different aminoacids 
in protein structures. We study,
by means of Monte-Carlo (MC) simulations, the 
dynamics of a 36-monomer chain 
characterized by a polymer sequence, 
denoted as $S_{36}$, designed by
minimizing the
energy in the target (native) 
conformation \cite{TRAPS} 
(see Fig. 1). 
In the units we are considering, 
the energy of $S_{36}$ in its
native structure is $E_{nat}=-16.5$, 
and the folding transition temperature  $T_f=0.40$. These values 
correspond to the 
normalized values $\tilde E_{nat}=-58.20$ and $\tilde T_f=1$ obtained 
for this sequence in 
Ref. \cite{TRAPS}. MC simulations 
of folding are performed using a
standard algorithm described extensively 
in the literature \cite{Algorithm},
in which, at each MC step, a monomer is picked up at random and corner flips 
and crankshaft moves are considered.

Mutations in the designed sequence are introduced by 
replacing a single monomer in $S_{36}$ by a monomer of a different type. In our case,
there exist 19 such possible substitutions for each monomer in $S_{36}$, implying
$36\times 19=684$ different sequences to study. 
Actually, a systematic behavior of folding can be deduced
by analyzing a relatively small number of such altered chains.
To show this, we started our analysis by choosing a monomer of $S_{36}$ arbitrarily, 
say the 17-th monomer, and studied the folding 
dynamics of the corresponding 19 altered chains.
The MC simulations were performed up to a maximum number of $15\cdot 10^6$ steps,
for the temperature $T=0.28$ and averaging over $12$ different starting
random coil configurations for each altered sequence.
By repeating the same procedure for other few selected sites along the original chain, 
it is possible to conclude that the behavior of altered chains 
can be generally classified into three categories: 

(1) chains which still fold to the native structure, 

(2) chains which fold to a unique compact structure, but different than the native one, and

(3) chains which, although becoming compact, do not fold to a unique structure at all
during simulations.

To characterize quantitatively the above 
three different behaviors, we find that
the quantity $\Delta E_{loc}$ defined as the 
difference between the energies of the altered 
and the intact (''wild-type'') chain, both calculated in 
the native configuration, plays a key role.
More precisely, such local energy 
difference, $\Delta E_{loc}[m'(i)\to m(i)]$,
for a mutation at site $i$, where the monomer 
$m(i)$ in $S_{36}$ is replaced by a monomer 
$m'(i)\ne m(i)$ is given by
\begin{equation}
\Delta E_{loc}[m'(i)\to m(i)]=\sum_{j\ne i}~(U_{m'(i),m(j)}-U_{m(i),m(j)})~
\Delta(\vert {\bf r}_i-{\bf r}_j\vert).
\label{deltaeloc}
\end{equation}

We have calculated all the 684 values of $\Delta E_{loc}[m'\to m]$, and found
they fall in the range $0\le \Delta E_{loc}[m'\to m]< 5.66$. 
We classify the impact of mutation by the ability of the mutated sequence
to fold into or close to the native conformation.
We define the degree of folding $Q$ similar to
earlier publications (see e.g. \cite{PRL1}) as the fraction of 
correctly formed contacts ($Q=1$ corresponds to the native state
and $Q \ll 1$ corresponds to misfolded states).

The following rules are obtained from our results:

(1) $\Delta E_{loc}[m'\to m]<1$: the altered chain always folds to the native structure
    ($Q=1$).

(2) $1<\Delta E_{loc}[m'\to m]<2.5$ 
: the altered chain folds to a unique structure, sometimes different than the native one, 
with $Q$ being smaller but close to one.
           In some cases, however, folding to the native structure may still occur ($Q=1$). 

(3) $2.5<\Delta E_{loc}[m'\to m]<4$: twilight zone: for some mutations, chains
fold into near native structure with $Q \sim 1$, other mutations
lead to misfolding with $Q \ll 1$.

(4) $\Delta E_{loc}[m'\to m]>4$: the altered chain does not fold to a unique
           structure at all during the simulation time, and now $Q \ll 1$.

For a given site $i$, we can therefore classify the 19 possible mutations
 according 
to the rules (1) and (3,4). For rule (2), additional information about the
 dynamical
behavior of the chain is required. 
We find that $73.675\%$ mutations fall into the first
class $\Delta E_{loc}[m'\to m]<1$, $26.3\%$ into the second class 
$1<\Delta E_{loc}[m'\to m]<4$,
and the rest $0.015\%$
into the class (4). Thus, a relatively large fraction of mutations
yield altered chains still folding into the native structure, some mutations
lead to limited misfolding ($Q \approx 1$) and only
a small fraction of mutations leads to complete misfolding.

Mutations at a given site may yield values $\Delta E_{loc}[m'\to m]$ which do not 
correspond to a single class, i.e. sometimes values smaller and larger than one
occur at the same site. However, assuming that mutations are not selective, i.e.
that they all occur with equal probability at a single site, an approximate 
scheme can be envisaged to classify the different sites according to the average
magnitude of the damage caused by mutations. This is done by
calculating the average value of $\Delta E_{loc}[m'\to m]$ for each site $i$ as, 
\begin{equation}
\Delta {\bar E}_{loc}(i)={1\over 19}\sum_{m'} \Delta E_{loc}[m'(i)\to m(i)].
\label{deltaaver}
\end{equation}
In this way, to each site $i$ is associated a mean value $\Delta {\bar E}_{loc}(i)$ 
and the following simple scheme emerges:

(1) When $\Delta {\bar E}_{loc}(i)<1$, chains having mutations at site $i$ are likely 
    to belong to the first category, i.e. on average they fold to the native structure
    and we denote $i$ as a 'cold' site,

(2) when $1<\Delta {\bar E}_{loc}(i)<2$ they behave on average as in the second category
    mentioned above, and $i$ is denoted as a 'warm' site. Finally,
    
(3) when $\Delta {\bar E}_{loc}(i)>2$ the resulting altered chains are likely to yield 
    unfolded structures, and the site is denoted as a 'hot' site.

We have classified the 36 monomers of $S_{36}$ according to this scheme as
shown schematically in Fig. 1. We find that 27 sites can be considered as cold sites,
6 as warm sites and only 3 as hot ones (see Table I). 
Thus, about 75\% of the heteropolymer chain
admits single error substitutions in the correct amino acid sequence yielding altered
chains still folding to the native structure. Only in a relatively small fraction 
of the chain (about 10\%), mutations have catastrophic effects leading to complete
misfolding. Additional simulations have confirmed the general trend predicted 
by this empirical scheme.

In order to gain deeper insight into 
the observed behavior of chains with mutations,
we evaluated the density of states $\nu (Q,E)$ which is the logarithm of 
the number of conformations having given values of the degree of folding
$Q$ and of the energy $E$, using histogram technique \cite{HISTOGRAM}. 
To this end long runs ($\approx 30\cdot 10^6$) were performed for the ''wild-type'' 
sequence and mutations S17L ($\Delta E_{loc}=1.12$), 
P16D ($\Delta E_{loc}=1.44$), D6E ($\Delta E_{loc}=0.44$)
and W16D ($\Delta E_{loc}=2.30$). 
In the latter case
the chain folded to a conformation, different from
the native one, having $Q=0.85$. 
(We used the following notation for mutations:
S17L means that at site 17 amino acid of type ``S'' is replaced
by amino acid of type ``L'').

Frequencies of appearance of conformations
with different values of $Q$ and $E$ were evaluated, and the analysis
described in  \cite{GAS95} was applied
to derive $\nu (Q,E)$. Fig. 2 shows the 3D plots for the density of states
for the ''wild-type'' sequences as well as  S17L for mutation. 
The remaining plots are qualitatively similar to the one
corresponding to mutation S17L and are not shown here to save space.
The pronounced boundary of ''continuous'' spectrum at $E_c \approx -14$
can be seen on all these plots. This is the lowest 
limiting energy which
non-native conformations (with $Q$ in the range $0,...,0.5$) have.
Comparing Fig. 2a with Fig. 2b,2c one can see that mutations 
do not affect $E_c$, i.e. the spectrum of energies of non-native
conformations. In contrary, the most pronounced differences between
the original sequence and mutants can be seen in the 
vicinity of the rightmost
of the low-energy ''tail'' corresponding to native and near-native 
conformations. One can define the physically meaningful 
''energy gap'' as the difference
in energy between the native state and lowest energy {\em misfolded}
(low $Q$) conformations i.e. $E_{nat}-E_c$.
Fig. 2 also helps to understand the difference between
the definition of the energy gap used in \cite{GAS95,SSK1} and the 
one used in a more recent work \cite{KT}.
Klimov and Thirumalai define ''energy gap'' as the difference between 
the energy of the native state conformation and nearest to it in energy.
It can be seen from Fig. 2 that the definition used in \cite{KT} 
concerns conformations nearest to the native in the high-$Q$ ''tail'',
i.e. differing from the native state by a monomer flip on the edge of
the structure.

Now, the results of our analysis can be easily rationalized.
The energy gap for the original sequence is -2.5 (i.e. $8k_{B}T$ at
the temperature of simulations). Mutations eliminating the
energy gap (having $\Delta E_{loc} > 2.5$) are in most cases disruptive,
so that mutant sequences lose all useful features of design,
behaving effectively as random ones, which, with some low probability, 
can still fold \cite{SSK1}. Mutations
with less pronounced energetic consequences
preserve the ''high-$Q$'' tail in histograms in Fig. 2 (sometimes
a new conformation from this tail may become native)
and sequences are still able to fold. These findings
once again emphasize the role of properly defined
''energy gap'' as an important factor determining folding properties.

The amino acid design tends to place
strongly interacting amino acids in internal
positions which form most contacts, creating therefore
a network of interactions inside the molecule 
which carries significant
part (up to $30\%$) of the total stabilization 
energy for the native structure. As a consequence protein structure 
is most sensitive to
mutations in these positions. Moreover, as 
was argued in Ref. \cite{CONSERV},
internal stabilizing amino acids are most 
likely to form their correct
contacts in the transition state, thus these residues can serve as
nucleation ones. 
For this reason, residues identified in \cite{NUCLEUS}
as nucleation sites are also ``hot'' positions as found
in the present study where mutations have most pronounced
impact on stability.

Our results provide additional insights
into the important issue of what part 
of the energy landscape of a protein is most affected
by mutations. We showed that the native and near-native
conformations are most affected while the impact of
mutations on the energies
of the ensemble of misfolded states are relatively less pronounced.
While this result differs from previous assertions \cite{Shortle} 
we believe that it makes clear physical sense. Indeed, in 
optimized sequences, each amino acid forms mostly favorable
contacts in the native state. Correspondingly, mutations
are doomed to replace many such interactions by less favorable
ones, significantly affecting the energy of the native state.
In contrast, in the unfolded state which represents an ensemble
of conformations, each amino acid forms flickering contacts with
many other amino acids and the impact of mutations
on the denatured state ''averages out'' over a multitude
of conformations of the denatured state ensemble. 
The assertion \cite{Shortle} that mutations
affect rather the denatured state was based on the analysis
of very short (16 monomers) 2-dimensional chains,
and the small size and limited conformational freedom
in two dimensions, is likely to have hindered the effect of averaging
over the conformational ensemble of the denatured state
which we observe in longer 3-dimensional chains.

An important issue is how general our results are,
in particular how do mutations affect folding of longer chains.
The energy gap (correctly defined, see Fig.2 and accompanying discussion) must be extensive in chain length
in order ensure fast folding to stable native conformation \cite{GSW,SG93}.
To this end longer chains should be more tolerant to point mutations
since they cause  smaller relative change of energy gap. However, 
still mutations at different positions will have 
different impact, and this factor may be crucial 
for longer chains as well
especially taking into account their multidomain behavior. 
In that case mutations will affect folding of a domain
to which mutated aminoacid belongs leaving 
stability of other domains unchanged
\cite{DOMAIN}.

Our study suggests that
protein structures may be remarkably 
tolerant to many mutations in a number of 'cold' positions.
Experimental studies \cite{Creighton} indicate
that this is indeed the case. It is likely 
that gradual changes may
accumulate to provide new proteins with different structure and 
function. 
We  believe that such evolutionary phenomenon can be observed
and studied in the realm of a simple folding model. 
It is the matter of future work justify or rule out this optimistic
assertion.

\bigskip
\bigskip
\bigskip
\bigskip
\noindent


\widetext 
\begin{table}
\caption{The average values of $\Delta {\bar E}_{loc}$ for each site of $S_{36}$. 
         Bold numbers indicate the hot sites.}

\begin{tabular}{cccccccc}
  \multicolumn{1}{c}{  site                   }&
  \multicolumn{1}{c}{$\Delta {\bar E}_{loc}$  }&
  \multicolumn{1}{c}{  site                   }&
  \multicolumn{1}{c}{$\Delta {\bar E}_{loc}$  }&
  \multicolumn{1}{c}{  site                   }&
  \multicolumn{1}{c}{$\Delta {\bar E}_{loc}$  }&
  \multicolumn{1}{c}{  site                   }&
  \multicolumn{1}{c}{$\Delta {\bar E}_{loc}$  }\\
\tableline
   1  &  0.24     &  10  &  0.64  &  19  &  0.38     &  28  &  1.31     \\
   2  &  0.73     &  11  &  1.46  &  20  &  0.63     &  29  &  0.95     \\ 
   3  &  1.94     &  12  &  0.50  &  21  &  0.65     &  30  &{\bf 2.48} \\
   4  &  0.75     &  13  &  0.55  &  22  &  0.40     &  31  &  0.99     \\
   5  &  1.09     &  14  &  1.85  &  23  &  0.28     &  32  &  0.78     \\
   6  &{\bf 2.79} &  15  &  0.88  &  24  &  0.77     &  33  &  0.68     \\
   7  &  0.77     &  16  &  1.79  &  25  &  0.44     &  34  &  0.42     \\
   8  &  0.27     &  17  &  0.80  &  26  &  0.54     &  35  &  0.66     \\
   9  &  0.09     &  18  &  0.38  &  27  &{\bf 3.46} &  36  &  0.26     \\
\end{tabular}
\label{table1}
\end{table}
\narrowtext


\begin{figure}
\caption{The native structure and sequence $S_{36}$ 
          of the model 36-mer.
          Schematic representation of the three types of sites 
          (different gray scale) characterizing single mutations in 
          the designed amino acid sequence according to the 
          value $\Delta {\bar E}_{loc}(i)$ (see text).
          With most  mutations occuring at a cold site (white balls), the chain
          can still fold to the native structure but the average folding time 
          is considerably longer than for the designed sequence.
          For most mutations occurring at a warm site (grey balls) the 
          chain folds to a unique structure, close to but 
          different than the 
          native one. For most mutations occurring at the hot sites (dark balls), 
          the chain does not fold to a unique structure, during simulation time   
          and different states of comparable local energy minima are 
          reached.}
\label{fig1}
\end{figure}

\begin{figure}
\caption{The density of states $\nu (Q,E)$ for: 
         (a) ``wild-type'' sequence, (b) mutation S17L} 
\label{fig2}
\end{figure}


\begin{thebibliography}{99}
\bibitem{Fersht}
L. Itzhaki, D. Otzen. and A. Fersht, J. Mol. Biol. {\bf 254}, 260 (1995).

\bibitem{NATUR}
E.I. Shakhnovich and A.M. Gutin, Nature {\bf 346}, 773 (1990).


\bibitem{SG93}
E.I. Shakhnovich and A.M. Gutin, Proc. Natl. Acad. Sci. USA {\bf 90}, 7195 (1993).


\bibitem{Shortle} D. Shortle, H.S. Chan and K. Dill, Protein Sci. {\bf 1}, 201 (1992).

\bibitem{Creighton} T.E. Creighton, {\em Protein Folding} (Freeman, San Francisco, 1992).

\bibitem{Lat_Mod} E.I. Shakhnovich, Phys. Rev. Lett. {\bf 72}, 3907 (1994); J. Maddox, Nature {\bf 370}, 13 (1994); H.~S. Chan and K.~A. Dill, J.~Chem.~Phys. {\bf 100}, 9238 (1994); N.~D. Socci and J.~N. Onuchic, J.~Chem.~Phys. {\bf 100}, 1519 (1994); M.-
H.Hao and H.Scheraga, J.Phys.Chem. {\bf 98}, 4940 (1994).





\bibitem{NUCLEUS}
V.I. Abkevich, A.M. Gutin, and E.I. Shakhnovich,  Biochemistry, {\bf 33}, 10026 (1994).

\bibitem{CONSERV} E.I. Shakhnovich, V.I. Abkevich and O.B. Ptitsyn, Nature, {\bf 379}, 96 (1996).

\bibitem{MJ}
S. Miyazawa and R.~L. Jernigan, Macromolecules {\bf 18}, 534 (1985).



\bibitem{TRAPS}
V.I. Abkevich, A.M. Gutin, and E.I. Shakhnovich, J. Chem. Phys. {\bf 101}, 5062 (1994).



\bibitem{Algorithm} P.H. Verdier, J. Chem. Phys. {\bf 59}, 6119 (1973);
                    H.J. Hilhorst and J.M. Deutch, J. Chem. Phys. {\bf 63}, 5153 (1975);
                    A. Sali, E.I. Shakhnovich and M. Karplus, J. Mol. Biol. {\bf 235}, 1614 (1994);  

\bibitem{PRL1}
E.I. Shakhnovich {\em et al} Phys. Rev. Lett. {\bf 67}, 1665 (1991);


\bibitem{HISTOGRAM} A.M. Ferrenberg and R.H. Swensen, Phys. Rev. Lett. {\bf 63}, 1195 (1989);
                    A. Sali, E.I. Shakhnovich, and M. Karplus, Nature  {\bf 369}, 248 (1994);
                    N. Socci and J. Onuchic, J. Chem. Phys. {\bf 103}, 4732 (1995).

\bibitem{GAS95} A.M. Gutin, V. Abkevich and E. Shakhnovich, Proc. Natl. Acad. Sci. USA {\bf 92}, 1282 (1995).
    
\bibitem{SSK1} A. Sali, E.I. Shakhnovich, and M. Karplus, J. Mol. Biol. {\bf 235}, 1614 (1994).

\bibitem{KT} D. Klimov and D. Thirumalai, Phys. Rev. Lett. {\bf 76}, 4070 (1996).

\bibitem{DOMAIN}  V.Abkevich,  A. Gutin, and E.Shakhnovich,
Protein Science {\bf 4}, 1167--1177
(1995)



\bibitem{GSW} R.Goldstein, Z.A. Luthey-Schulten, and P.Wolynes.
 Proc. Natl. Acad. Sci. USA {\bf 89}, 4918 (1992).



 
\end{thebibliography}
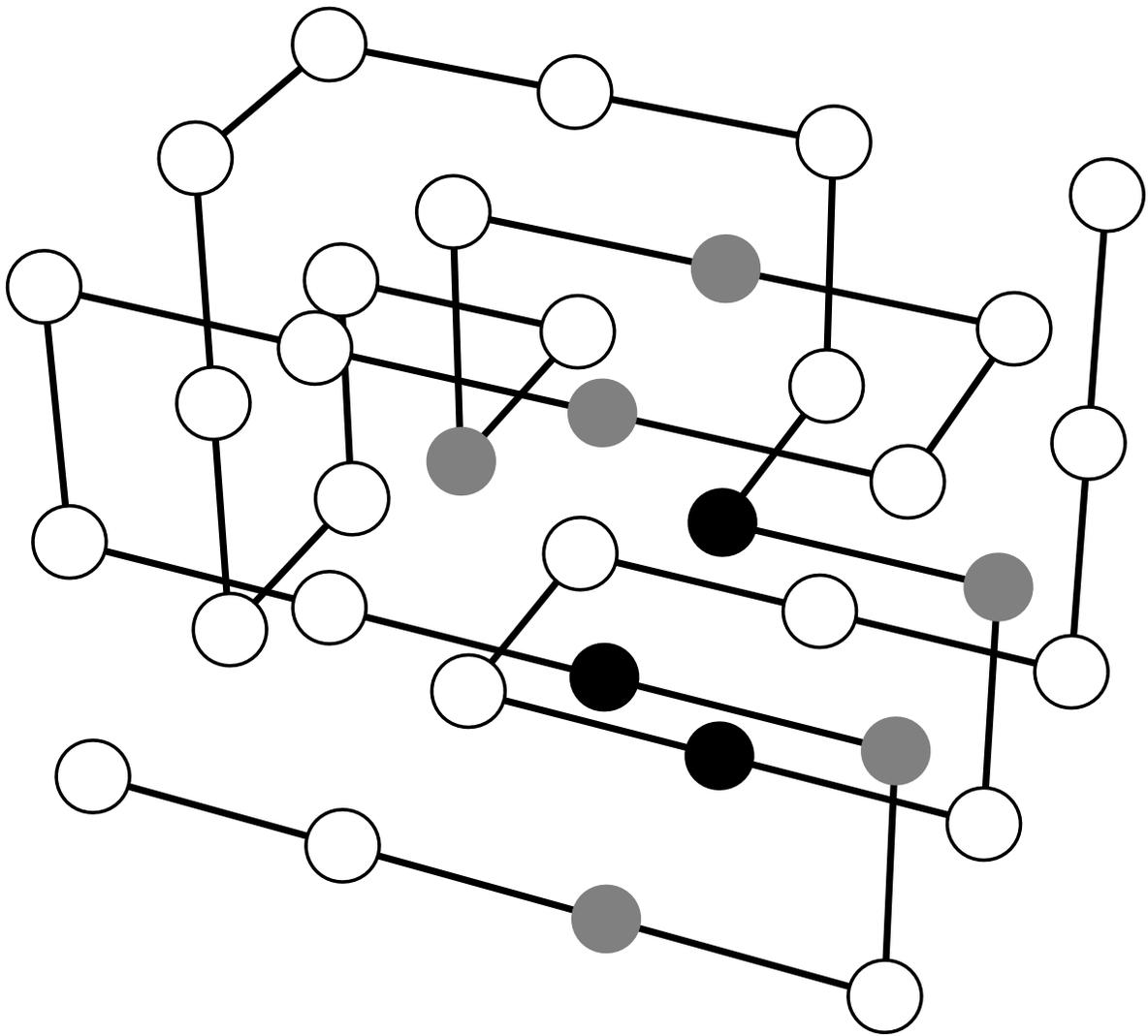
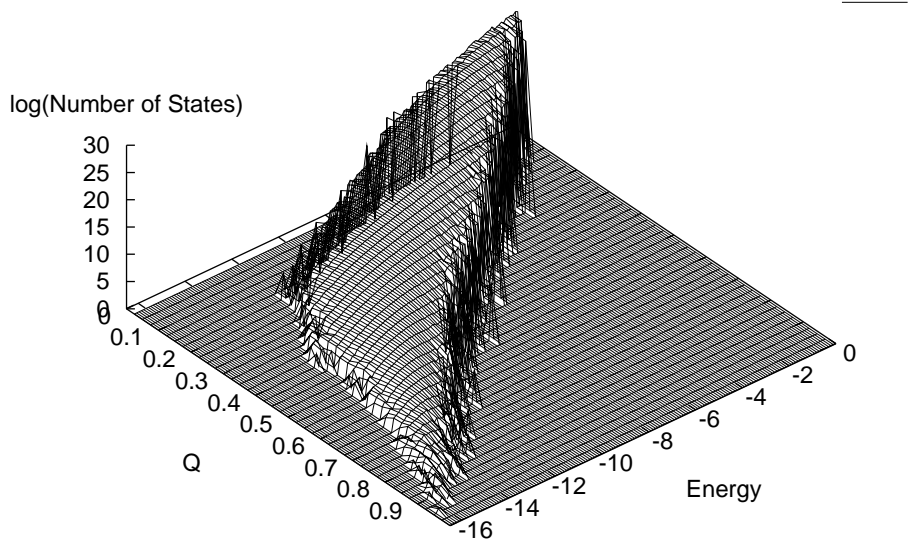
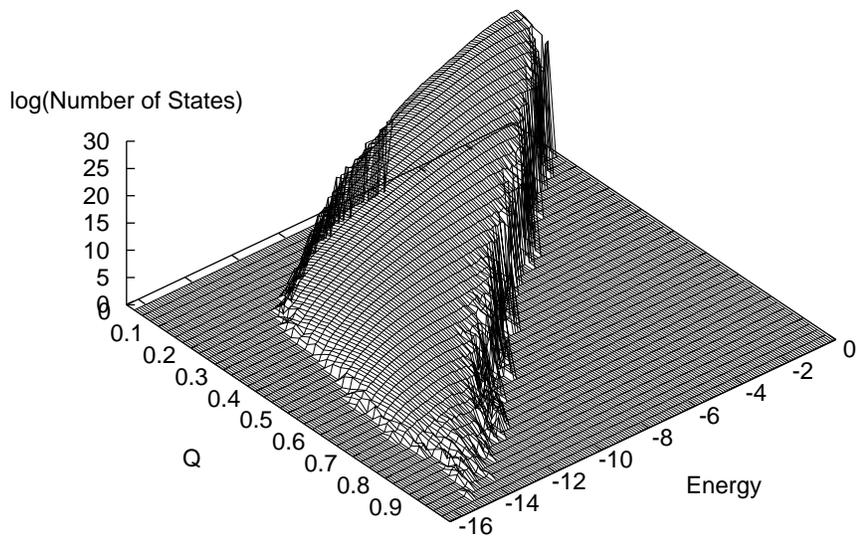
\end{document}